\documentclass[11pt]{article}
\usepackage{amssymb,amsmath,cite,geometry}
\catcode`\@=11

\@addtoreset{equation}{section}
\def\theequation{\arabic{section}.\arabic{equation}}
     
\catcode`\@=11
\def\thesection{\arabic{section}}

\def\appendix{\setcounter{section}{0}
        \def\thesection{Appendix.}
        \def\theequation{\Alph{section}.\arabic{equation}}}
\def\section{\@startsection{section}{1}{\z@}{3.5ex plus 1ex minus
  .2ex}{2.3ex plus .2ex}{\large\bf}}
\makeatletter
\renewcommand{\@seccntformat}[1]{\csname the#1\endcsname.\quad}
\makeatother

\long\def\@makefntext#1{\parindent 0cm\noindent
\hbox to 1em{\hss$^{\@thefnmark}$}#1}

\newcommand{\captionfonts}{\small}
\makeatletter  
\long\def\@makecaption#1#2{%
  \vskip\abovecaptionskip
  \sbox\@tempboxa{{\captionfonts #1: #2}}%
  \ifdim \wd\@tempboxa >\hsize
    {\captionfonts #1: #2\par}
  \else
    \hbox to\hsize{\hfil\box\@tempboxa\hfil}%
  \fi
  \vskip\belowcaptionskip}
\makeatother   

\begin{document}
\begin{titlepage}
\vspace{.5in}
\begin{flushright}
March 2015\\ 
revised July 2015 
\end{flushright}
\vspace{.5in}
\begin{center}
{\Large\bf
 Four-Dimensional Entropy from Three-Dimensional Gravity}\\  
\vspace{.4in}
{S.~C{\sc arlip}\footnote{\it email: carlip@physics.ucdavis.edu}\\
       {\small\it Department of Physics}\\
       {\small\it University of California}\\
       {\small\it Davis, CA 95616}\\{\small\it USA}}
\end{center}

\vspace{.5in}
\begin{center}
{\large\bf Abstract}
\end{center}
\begin{center}
\begin{minipage}{4.95in}
{\small
At the horizon of a black hole, the action of (3+1)-dimensional loop
quantum gravity acquires a boundary term that is formally identical
to an action for three-dimensional gravity.  I show how to use this 
correspondence to obtain the entropy of the (3+1)-dimensional black 
hole from well-understood conformal field theory computations of 
the entropy in (2+1)-dimensional de Sitter space.
}
\end{minipage}
\end{center}
\end{titlepage}
\addtocounter{footnote}{-1}

The ability to explain black hole thermodynamics is a key test of any 
quantum theory of gravity.  In this regard, loop quantum gravity has
a mixed record.  The correct area dependence of black hole entropy  
appears quite naturally \cite{ABCK,ABK}.  But to obtain quantitative 
agreement with the semiclassical results of Bekenstein 
and Hawking, it seems necessary to tune a rather mysterious 
parameter, the Barbero-Immirzi parameter $\gamma$, to a peculiar 
value determined by a complex combinatorial computation 
\cite{Domagala,Meissner}.

In the past few years, there have been intriguing hints that the 
entropy can also be obtained by setting $\gamma=i$ 
\cite{Geiller,Achour,GhoshPran,Carlipz}.  This is the natural value:
it makes the theory self-dual \cite{Ashtekar}, and is the 
only choice for which the Ashtekar-Barbero-Sen connection (\ref{a1})
is a fully diffeomorphism-invariant spacetime connection 
\cite{Samuel,Alexa}.  Unfortunately, with this choice one must
impose a reality conditions, a procedure that remains poorly
defined.  As a consequence, the theory with $\gamma=i$ is not 
nearly as mathematically sophisticated as the version with real 
$\gamma$, and far fewer results have been established.

In this paper, I will describe a simple new method for computing 
black hole entropy in loop quantum gravity with $\gamma=i$.  
The key observation is that loop quantum gravity requires a 
boundary term at a black hole horizon that is formally identical 
to an action for three-dimensional gravity with a positive 
cosmological constant.  The identification is not an obvious 
geometric one, but the four-dimensional horizon maps to a 
well understood three-dimensional spacetime, and 
one can exploit this association to use standard techniques 
of conformal field theory to count the states.  

\section{Two $\hbox{SL}(2,\mathbb{C})$ actions}

We start with  (3+1)-dimensional gravity in first-order form, 
treating the tetrad one-form $e^I = e_\mu{}^Idx^\mu$ and 
the spin connection one-form $\omega^{IJ} =\omega_\mu{}^{IJ}dx^\mu$ 
as independent variables.   The Ashtekar-Sen self-dual connection 
\cite{Ashtekar,Sen} is $A^{IJ} = \frac{1}{2}%
\left(\omega^{IJ} + \frac{i}{2}\epsilon^{IJ}{}_{KL}\omega^{KL} \right)$,
but to avoid double-counting components, it is sufficient to consider
the complexified $\hbox{SU}(2)$---or equivalently, 
$\hbox{SL}(2,\mathbb{C})$---connection 
\begin{align}
A^i = i\omega^{0i} + \frac{1}{2}\epsilon^{ijk}\omega_{jk} ,
\label{a1}
\end{align}
where lower case Roman indices run from $1$ to $3$ (see, for instance, 
section 4.3 of \cite{Rovelli}).   The gravitational action can then be written 
in the form \cite{JacobsonSmolin,Samuelb}
\begin{align}
I_4 = -\frac{i}{16\pi G_4}\int\! d^4x\, \Sigma_i\wedge F^i ,
\label{a2}
\end{align}
where $F^i = dA^i + \epsilon^{ijk}A_j\wedge A_k$ is the curvature 
of the connection and 
$\Sigma^i = ie^0\wedge e^i + \frac{1}{2}\epsilon^{ijk}e_j\wedge e_k$
is the self-dual projection of $e^I\wedge e^J$.
The real part of (\ref{a2}) is equal to the standard
Einstein-Hilbert action, while the imaginary part is essentially
irrelevant: it is extremal whenever the real part is, so it does not
change the equations of motion, and it vanishes on shell. 
In loop quantum gravity, the factor of $i$ in (\ref{a1}) is often replaced
by an arbitrary parameter $\gamma$, the Barbero-Immirzi parameter.
The quantization becomes much simpler when $\gamma$ is chosen to
be real, but as noted above, hints are now appearing that the self-dual 
choice $\gamma=\pm i$ simplifies and clarifies the description of
black hole entropy.

Now suppose that a black hole is present, with a horizon $\Delta$
of area $A_\Delta$.  For the surface $\Delta$ to be an isolated horizon 
\cite{ISO}, it must obey a geometric restriction, which translates to 
the condition \cite{ABCK,Krasnova}
\begin{align}
F^i = -\frac{2\pi}{A_\Delta}\Sigma^i \quad\hbox{on $\Delta$.} 
\label{a5}
\end{align}  Although the horizon is not a physical boundary, the
imposition of (\ref{a5}) forces us to add a ``boundary'' term to the
action.  As first noted by Smolin in a slightly different context 
\cite{Smolin}, the required term is a Chern-Simons action.  The
specific form depends on the Barbero-Immirzi parameter; for
our choice $\gamma = i$, it is a chiral $\hbox{SL}(2,\mathbb{C})$ 
Chern-Simons action
\begin{align}
I_\Delta = \frac{k}{4\pi} \int_\Delta \mathrm{Tr}
  \left\{A\wedge dA + \frac{2}{3} A\wedge A\wedge A \right\} ,
\label{a6}
\end{align}
where $A = A^iT_i$ is the $\hbox{sl}(2,\mathbb{C})$-valued connection
with generators normalized so $\mathrm{Tr}(T_iT_j) = \frac{1}{2}\eta_{ij}$,
and the coupling constant $k$ is expressed in terms of (3+1)-dimensional 
gravitational quantities as
\begin{align}
k_{4D} = \frac{iA_\Delta}{8\pi G_4} .
\label{a7}
\end{align}
Moreover, the symplectic form---that is, the set of Poisson brackets%
---also acquires a boundary term for the connection at the horizon, 
which is identical to the symplectic form of Chern-Simons theory (see, e.g., 
\cite{Pranzetti}).  Thus components of the connection, which commute
in the bulk, become canonically conjugate at $\Delta$, and by the usual 
rules of quantization we expect a Hilbert space 
$\mathcal{H}_{\mathrm\scriptstyle bulk}\otimes \mathcal{H}_\Delta$, 
with the bulk and horizon states related by the operator version of the 
boundary conditions (\ref{a5}) \cite{ABK}.

So far, I have not used loop quantum gravity.  I now exploit 
one general feature of that quantization.  Classically, the boundary 
conditions (\ref{a5}) imply that the boundary $\hbox{SL}(2,\mathbb{C})$ 
connection is not flat, and is thus not an extremum of the Chern-Simons 
action.  In loop quantum gravity, though, quantum states are described 
by spin networks, and the area element on the right-hand side of 
(\ref{a5}) is distributional, differing from zero only at the ``punctures'' 
where spin network edges intersect the horizon.  The boundary conditions
\emph{are} then equivalent to the equations of motion for a Chern-Simons
theory, but now on a sphere with punctures (or, technically, a manifold
$\mathbb{R}\times S^2$ with Wilson lines) \cite{WittenCS}.  Hence the 
boundary Hilbert space $\mathcal{H}_\Delta$ is that of a Chern-Simons 
theory on a sphere with punctures.   In standard loop quantum gravity, 
one can say much more---holonomies around punctures give calculable 
elements of area---but we shall not need any of those details; it is
enough that the boundary theory acts as an independent Chern-Simons
theory coupled to the bulk through a set of punctures.

The action (\ref{a6}) also appears in a very different context, though:
it is the first-order action for (2+1)-dimensional gravity with
a positive cosmological constant $\Lambda=1/\ell^2$ \cite{Wittenx}.  
The connection is now 
\begin{align}
{\tilde A}^a = \frac{1}{2}\epsilon^{abc}{\tilde\omega}_{bc} 
     + \frac{i}{\ell}{\tilde e}^a  ,
\label{a8}
\end{align}
where ${\tilde e}^a$ and ${\tilde\omega}^{bc}$ are the three-dimensional
triad and spin connection, and the coupling constant $k$ is  
\begin{align}
k_{3D} = \frac{i\ell}{2G_3}  ,
\label{a9}
\end{align}
now expressed in terms of (2+1)-dimensional quantities.
Much as in the four-dimensional case, the real part of (\ref{a6}) gives
the usual Einstein-Hilbert action, while the imaginary part is an
``exotic'' term that is extremal when the real part is extremal and
vanishes on shell.\footnote{The (2+1)-dimensional 
action is usually written as a 
sum of (\ref{a8}) and its complex conjugate, in which case the
coupling constant for each term is half of (\ref{a9}).  Here, though,
we wish to match the (3+1)-dimensional action, which is chiral.}
The $\hbox{SL}(2,\mathbb{C})$ Chern-Simons action is also related
to ``Euclidean anti-de Sitter space''; I will return to this point in the
conclusion.

Although the two appearances of the Chern-Simons action both
involve gravity, their mathematical equivalence is, as far as I know,
purely accidental.  One can try to construct a geometrical relationship,
but if one exists, it is subtle.  Indeed, comparing (\ref{a1}) and 
(\ref{a8}), we see that while the connections can be made to match, 
the triad ${\tilde e}^a$ in three dimensions corresponds to the 
extrinsic curvature in four dimensions.  Hence we might not expect 
the three-dimensional theory to give a simple geometrical picture 
of the states (although see \cite{Carlipz,CarlipGF}).  Still, the formal 
equivalence of the actions will be enough to determine the 
(3+1)-dimensional Bekenstein-Hawking entropy.

\section{Entropy}

Let us focus for now on (\ref{a6}) as an action for (2+1)-dimensional
gravity.  For the case of a \emph{negative} cosmological constant,
the counting of states in such a theory is well understood 
\cite{Strominger,BSS}, although the exact nature of those states
is not \cite{Carlipbh}.  As Brown and Henneaux showed, the
asymptotic symmetry in such a theory is a two-dimensional
conformal symmetry \cite{BrownHenneaux}, which is powerful 
enough that the Cardy formula determines the asymptotic density 
of states without requiring any further details \cite{Cardy,Cardy2,Carlipc}.

For the case of a positive cosmological constant, there is no
asymptotic spatial boundary, and the picture is not as
clean.  One can, however, look at the asymptotic symmetries 
at timelike infinity \cite{Bala,Maloney,Umetsu}; or impose boundary
conditions on a tube, which can be viewed as the world line of an 
observer \cite{Park}; or continue to negative $\Lambda$ 
\cite{BBO}; or perhaps obtain a central charge directly from the
symmetries of the phase space \cite{Kelly}.
One obtains a consistent answer: a ``puncture'' with 
$\hbox{SL}(2,\mathbb{C})$ holonomy conjugate to
\begin{align}
H = \left(\begin{array}{cc} e^{\pi i r_+/\ell}&a\\0&e^{-\pi i r_+/\ell}
   \end{array}\right)
\label{b1}
\end{align}
gives a local geometry of a cone, and contributes an entropy 
\begin{align}
S = \frac{2\pi r_+}{4G_3} = -ik_{3D}\frac{\pi r_+}{\ell}  .
\label{b2}
\end{align}
(For subtleties coming from the fact that we are considering a
purely chiral action, see \cite{Parkx}.)

While (\ref{b2}) was derived in $2+1$ dimensions, it is ultimately 
a statement about the quantum mechanics of the action (\ref{a6}),
which is also the boundary action in $3+1$ dimensions.
We now use a single fact from the four-dimensional picture: a
cross-section of the horizon $\Delta$ at a fixed time is a two-sphere
$S^2$.  Consider a loop on this two-sphere surrounding all of
the punctures.  On the one hand, the $\hbox{SL}(2,\mathbb{C})$
holonomy of this loop is the product of the holonomies around
each puncture.  On the other hand, the loop also surrounds a
region with no punctures, for which the holonomy must be the
identity.  Assuming that all of the holonomies are in the same
conjugacy class (\ref{b1})---I will return to this below---it is easy
to see that this requires that
\begin{align}
\sum_{\hbox{\tiny\it punctures}} \frac{\pi r_+}{\ell} = 2\pi  .
\label{b3}
\end{align}
Thus from (\ref{b2}) and (\ref{a7}),
\begin{align}
S = -2\pi ik_{3D} = -2\pi ik_{4D} = \frac{A_\Delta}{4G_4}   ,
\label{b4}
\end{align}
reproducing the correct Bekenstein-Hawking entropy for the 
\emph{four}-dimensional black hole.

\section{The Schwarzschild black hole}

To make the discussion more concrete, let us specialize to the 
Schwarzschild black hole.  Following Kaul and Majumdar \cite{Kaul},
we write the metric in Kruskal-Szekeres form as
\begin{align}
ds^2 = -2B(r)dvdw + r^2(v,w)(d\theta^2 + \sin^2\theta d\varphi^2)
\quad\ \hbox{with $B = \frac{4r_+{}^3}{r\ }e^{-r/r_+}$,
$-2vw = \left(\frac{r\ }{r_+}-1\right)e^{r/r_+}$}
\label{c1}
\end{align}
and choose a tetrad
\begin{align}
e^0 = \sqrt{\frac{B}{2}}\left(\frac{w}{\alpha}dv + \frac{\alpha}{w}dw\right) ,\qquad
e^1 = \sqrt{\frac{B}{2}}\left(\frac{w}{\alpha}dv - \frac{\alpha}{w}dw\right) ,\qquad
e^2 = rd\theta ,\qquad 
e^3 = r\sin\theta d\varphi ,
\label{c2}
\end{align}
where $\alpha$ is an arbitrary function labeling a gauge choice for
local Lorentz transformations.  It is then straightforward to compute the
connection (\ref{a1}); at the horizon $r=r_+$, $B=B_+$, $w=0$, one 
finds \cite{Kaul}
\begin{align}
A^1 = \cos\theta d\varphi + i\frac{d\alpha}{\alpha} ,\ \
A^2 = -\sqrt{\frac{B_+}{2}}\frac{1}{2r_+}\alpha%
       \left(id\theta - \sin\theta d\varphi\right),\ \
A^3  = -\sqrt{\frac{B_+}{2}}\frac{1}{2r_+}\alpha%
       \left(i\sin\theta d\varphi  + d\theta\right)  .
\label{c3}
\end{align}

Now, by (\ref{a8}),  the imaginary part of the connection (\ref{c3}) should give 
the triad in the (2+1)-dimensional picture.  Defining
\begin{align}
\frac{1}{\beta} = \sqrt{\frac{B_+}{2}}\frac{1}{2r_+}\alpha
\label{c4}
\end{align}
we see that the classical (2+1)-dimensional metric is
\begin{align}
ds^2 = \frac{\ell^2}{\beta^2}\left( -d\beta^2 + d\theta^2 
        + \sin^2\theta d\varphi^2\right)  .
\label{c5}
\end{align}
This is \emph{almost} the de Sitter metric on an expanding patch.  It
is not quite; the curvature is not constant, but satisfies an equivalent of 
(\ref{a5}).  But as noted above, in loop quantum gravity we should 
replace the continuous curvature by a collection of punctures, of the 
type first introduced by Deser and Jackiw \cite{Deser}.  That is, as
in Regge calculus, we should replace (\ref{c5}) by a locally de Sitter metric 
\begin{align}
ds^2 = \frac{\ell^2}{\beta^2}\left( -d\beta^2 + dzd{\bar z}\right)
\quad\hbox{with $z=x+iy$}
\label{c6}
\end{align}
with a set of conical singularities that reproduce the curvature of (\ref{c5})
in the large. 

Now, the isometry group of the de Sitter metric (\ref{c6}) is 
$\hbox{SL}(2,\mathbb{C})$, with an action \cite{sl}
\begin{align}
\left(\begin{array}{cc} a&b\\c&d\end{array} \right): (z, \beta) \rightarrow
\left(\frac{(az+b)({\bar c}{\bar z} + {\bar d}) + a{\bar c}\beta^2}%
{|cz+d|^2 + |c|^2\beta^2}, \frac{\beta}{|cz+d|^2 + |c|^2\beta^2}\right)  .
\label{c7}
\end{align}
To obtain the metric (\ref{c5}) from (\ref{c6}), we must add a set of conical
points on surfaces of constant $\beta$.  The condition for an isometry
(\ref{c7}) to preserve such surfaces is that $c=0$, $|d|=1$, and
the resulting isometries are precisely the ones given by (\ref{b1}).

In slightly more detail, an elliptic element
\begin{align}
R = \left(\begin{array}{cc} e^{i\theta}&0\\0&e^{-i\theta}\end{array} \right)
\label{c8}
\end{align}
of $\hbox{SL}(2,\mathbb{C})$ rotates $z$ by an angle $2\theta$ around 
the origin, while a parabolic element
\begin{align}
T = \left(\begin{array}{cc} 1&a\\0&1\end{array} \right)
\label{c9}
\end{align}
translates $z$ by $a$.  An individual puncture at position $a$ thus 
corresponds to a holonomy $TRT^{-1}$, equivalent to (\ref{b1}), and 
the total holonomy is
\begin{align}
H = T_1R_1T_1{}^{-1}T_2R_2T_2{}^{-1}T_3R_3T_3{}^{-1}\dots
\label{c10}
\end{align}
in agreement with the analysis of the preceding section.

We can now go further.  The parabolic element $T$ can be written as
\begin{align}
\left(\begin{array}{cc} 1&a\\0&1\end{array} \right) = \exp\left\{ a(J_1 + K_2)\right\} ,
\label{c12}
\end{align}
where the generators of complexified $\hbox{SU}(2)$ are $J_i=\frac{1}{2}\sigma_i$,
$K_i =\frac{i}{2}\sigma_i$.  From a (3+1)-dimensional viewpoint, this is a
null rotation, a Lorentz transformation that leaves a null vector fixed.
Similarly, $RTR^{-1}$ is a null rotation fixing a different, rotated null vector.
The holonomy (\ref{c10}) can be rewritten as
\begin{align}
H = \bigl(T_1\bigr)\bigl(R_1(T_1{}^{-1}T_2)R_1{}^{-1}\bigr)
       \bigl((R_1R_2)(T_2{}^{-1}T_3)(R_1R_2)^{-1}\bigr)\dots
\label{c11}
\end{align}
that is, as a product of null rotations.   

This is just what one would expect in the self-dual formulation of general 
relativity, where the connection (\ref{a1}) involves a sum of a rotation
and a boost.  But we can now even identify the null vector being
held fixed.  The coordinate $\beta$ in the (2+1)-dimensional 
metric (\ref{c5}) originated as a gauge-dependent parameter in the 
(3+1)-dimensional tetrad (\ref{c2}).  But for (\ref{c5}) to be truly
(2+1)-dimensional, $\beta$ cannot depend on $\theta$ and $\varphi$
alone, but must be a function of the null coordinate $v$ along the horizon.  
Indeed,  to preserve spherical symmetry, $\beta$ should be a function of 
$v$ alone.  Hence the isometries (\ref{b1}), chosen in 2+1 dimensions to
leave $\beta$ invariant, fix $v$ in 3+1 dimensions.  The null  vector that
defines our null rotations is just the null normal to the horizon.   

This choice is physically natural, and may offer insights into the underlying
degrees of freedom \cite{Carlipz}.  But it is awkward to implement in a
formulation with a real Barbero-Immirzi parameter, perhaps explaining
why the derivation of black hole entropy is simpler with a self-dual
connection.

\section{Implications and open questions}

I have focused on the Schwarzschild black hole, but the general
arguments about the structure of holonomies hold for any black
hole satisfying the isolated horizon boundary conditions (\ref{a5}).  
Still, it would be interesting to see an explicit extension to an arbitrary
black hole.  For the Kerr black hole, much of the preliminary work
appears in \cite{Roken}, although a more general Lorentz gauge is
needed.

The present derivation of black hole entropy differs from the standard
loop quantum gravity approach of \cite{ABCK,ABK} in an interesting way.
The usual starting point is an ensemble of horizon configurations with
arbitrarily many punctures and arbitrary holonomies, restricted
only by the specified area $A_\Delta$.  Counting
states is then a combinatorial problem; an entropy proportional to
area appears naturally, but the Barbero-Immirzi parameter must be
tuned to give the right prefactor.  Here, in contrast, the entropy is
derived for a \emph{single} configuration of punctures and holonomies,
now restricted only by the closure condition (\ref{b3}).  This is 
reminiscent of the proposal that the number of punctures should 
be treated as a sort of ``quantum hair'' \cite{GhoshPerez} that
physically distinguishes different black holes.  In essence, the
question is in how fine a coarse-graining is needed to define the 
entropy.

The method of counting states here also differs from the standard
approach.  In contrast to the usual procedure, our central result 
(\ref{b2}) depends on no details of the Hilbert space, but only on 
the fact that an $\hbox{SL}(2,\mathbb{C})$ Chern-Simons theory 
implies a two-dimensional conformal symmetry, which is powerful 
enough to severely constrain the density of states.  Similar symmetry 
arguments have been used in other attempts to count black hole 
states---see \cite{CEntropy} for a review---and it is intriguing that 
the central charge $c=6k$ here is nearly identical to the value obtained 
in those approaches, differing by a factor of two.  Any relationship 
between these analyses must be a bit subtle, since the conformal
methods of \cite{CEntropy} involve symmetries in the ``$r$--$t$
plane'' rather than symmetries of spatial sections of the horizon.  But as 
Pranzetti has pointed out \cite{Pranzettib}, the self-dual connection 
(\ref{a1}) automatically links transformations in these different spaces,
so a relationship might exist.

There is another direction in which this work might be extended.  
An $\hbox{SL}(2,\mathbb{C})$ Chern-Simons theory is a theory of 
(2+1)-dimensional de Sitter gravity, but also of ``Euclidean anti-de 
Sitter gravity,'' that is,  (2+1)-dimensional gravity with $\Lambda<0$
analytically continued to Riemannian signature.  Punctures  then 
correspond to point particles in AdS, and the quantization is almost
 certainly related to Liouville theory \cite{Krasnov}.  An interesting 
new possibility now arises if we allow the elliptic holonomies (\ref{b1}) 
to lie in different conjugacy classes.  The product of a large number 
of random elliptic elements of $\hbox{SL}(2,\mathbb{C})$ is exponentially 
likely to be hyperbolic \cite{Furstenberg,Quint}, and a hyperbolic 
isometry in AdS signals the appearance of a three-dimensional black 
hole horizon.  It is not entirely clear how to count the resulting degrees 
of freedom---I do not know the analog of the closure condition (\ref{b3})%
---but this would be interesting to pursue.  One possibility is to use 
the canonical version of the Cardy formula; as discussed in section 5 
of \cite{Carlipz}, this may again yield the correct entropy.

Finally, it is interesting to ask whether this sort of calculation 
can be applied in more general contexts.  Boundary conditions of 
the form (\ref{a5}) have been suggested in broader settings \cite{Smolin}; 
it would be good to know whether the results of this paper can be
extended to, for instance, general spatial boundaries, or perhaps
causal horizons \cite{Jacobson}.  Note also that the derivation here
made very limited use of the dynamics of general relativity \cite{Husain};
while the results almost certainly depend on the use of a noncompact
connection \cite{Carlipz}, they might generalize to BF theory or its 
deformations \cite{KrasnovBF}.

\vspace{1ex}
\begin{flushleft}
\large\bf Acknowledgments
\end{flushleft}

I would like to thank Abhay Ashtekar, Marc Geiller, Kirill Krasnov, 
Daniele Pranzetti, and Jean-Fran{\c{c}}ois Quint for very helpful comments 
and conversations.
This work was supported in part by Department of Energy grant
DE-FG02-91ER40674.


\begin{thebibliography}{99}\addtolength{\itemsep}{-.8ex}
\bibitem{ABCK} A.\ Ashtekar, J.~C.\ Baez, A.\ Corichi, and K.\ Krasnov,
Quantum geometry and black hole entropy, {Phys.\ Rev.\ Lett.}
{80} (1998) 904, arXiv:gr-qc/9710007.

\bibitem{ABK} A.\ Ashtekar, J.~C.\ Baez, and K.\ Krasnov, 
Quantum geometry of isolated horizons and black hole entropy,
{Adv.\ Theor.\ Math.\ Phys.} {4} (2000) 1, 
arXiv:gr-qc/0005126.

\bibitem{Domagala} M.\ Domagala and J.\ Lewandowski, Black hole 
entropy from quantum geometry, {Class.\ Quant.\  Grav.}
{21} (2004) 5233, arXiv:gr-qc/0407051.

\bibitem{Meissner} K.~A.\ Meissner, Black hole entropy in loop 
quantum gravity, {Class.\ Quant.\  Grav.} {21}
(2004)  5245, arXiv:gr-qc/0407052. 

\bibitem{Geiller} E.\ Frodden, M.\ Geiller, K.\ Noui, and A.\ Perez,
Black hole entropy from complex Ashtekar variables, {Europhys.\
Lett.} {107} (2014) 10005, arXiv:1212.4060.

\bibitem{Achour} J.~B.\ Achour, A.\ Mouchet, and K.\ Noui,
Analytic continuation of black hole entropy in loop quantum 
gravity, arXiv:1406.6021. 

\bibitem{GhoshPran} A.\ Ghosh and D.\ Pranzetti, CFT/gravity 
correspondence on the isolated horizon, {Nucl.\  Phys.} {B889} 
(2014) 1, arXiv:1405.7056.

\bibitem{Carlipz} S.\ Carlip, A note on black hole entropy in loop 
quantum gravity, arXiv:1410.5763.

\bibitem{Ashtekar} A.\ Ashtekar, New variables for classical and 
quantum gravity, {Phys.\ Rev. Lett.} {57}  (1986), 2244.

\bibitem{Samuel} J.\ Samuel, Is Barbero's Hamiltonian formulation 
a gauge theory of Lorentzian gravity?, {Class.\ Quant.\  Grav.}
{17} (2000) L141, arXiv:gr-qc/0005095.

\bibitem{Alexa} S.\ Alexandrov, On choice of connection in loop 
quantum gravity, {Phys.\ Rev.\ D} {65} (2002) 024011,
arXiv:gr-qc/0107071.

\bibitem{Sen} A.\ Sen, Gravity as a spin system, {Phys.\ Lett.} {B 119}
(1982) 89.

\bibitem{Rovelli} C.\ Rovelli, \emph{Quantum Gravity} (Cambridge 
University Press, Cambridge, 2004).

\bibitem{JacobsonSmolin} T.\ Jacobson and L.\ Smolin, Covariant 
action for Ashtekar's form of canonical gravity, {Class.\ Quant.\ Grav.} 
{5} (1988) 583.

\bibitem{Samuelb} J.\ Samuel, A Lagrangian basis for Ashtekar's 
reformulation of canonical gravity, {Pramana} {28} (1987) L429.

\bibitem{ISO} A.\ Ashtekar,  A.\ Corichi, and K.\ Krasnov,
Isolated horizons: the classical phase space, {Adv.\ Theor.\ Math.\
Phys.} {3} (1999) 419, arXiv:gr-qc/9905089.

\bibitem{Krasnova} K.\ Krasnov, On quantum statistical mechanics 
of a Schwarzschild black hole, {Gen.\ Rel.\ Grav.} {30} (1998) 53,
arXiv:gr-qc/9605047.

\bibitem{Smolin} L.\ Smolin, Linking topological quantum field theory 
and nonperturbative quantum gravity, {J.\ Math.\ Phys.} {36} (1995) 
6417, arXiv:gr-qc/9505028.

\bibitem{Pranzetti} J.\ Engle, K.\ Noui, A.\ Perez, and D.\ Pranzetti,
Black hole entropy from an SU(2)-invariant formulation of Type I 
isolated horizons, {Phys.\ Rev.} {D82} (2010) 044050, arXiv:1006.0634.

\bibitem{WittenCS} E.\ Witten, Quantum field theory and the Jones 
polynomial, {Commun.\ Math.\ Phys.} {121} (1989) 351.

\bibitem{Wittenx} E.\ Witten, 2+1 dimensional gravity as an exactly
soluble system, {Nucl.\ Phys.\ B} 311 (1998) 46.

\bibitem{CarlipGF} S.\ Carlip, Symmetries, horizons, and black hole
entropy, {Gen.\ Rel.\ Grav.} {39} (2007) 1519,  arXiv:0705.3024.

\bibitem{Strominger} A.\ Strominger, Black hole entropy from near 
horizon microstates, {JHEP} {02} (1998) 009, 
arXiv:hep-th/9712251. 

\bibitem{BSS} D.\ Birmingham, I.\ Sachs, and S.\ Sen, Entropy of 
three-dimensional black holes in string theory, {Phys.\ Lett.}
{B424} (1998) 275, arXiv:hep-th/9801019.

\bibitem{Carlipbh} S.\ Carlip, Conformal field theory, (2+1)-dimensional 
gravity, and the BTZ black hole, {Class.\ Quant.\ Grav.} {22} (2005)
R85, arXiv:gr-qc/0503022.

\bibitem{BrownHenneaux} J.~D.\ Brown and M.\ Henneaux, Central 
charges in the canonical realization of asymptotic symmetries: an 
example from three-dimensional gravity, {Commun.\ Math.\ Phys.} 
{104} (1986) 207.

\bibitem{Cardy} J.~A.\ Cardy, Operator content of two-dimensional 
conformally invariant theories, {Nucl.\ Phys.\ B} {270}
(1986) 186.

\bibitem{Cardy2}  H.~W.~J.\ Bl{\"o}te, J.~A.\ Cardy, and M.~P.\ 
Nightingale, Conformal invariance, the central charge, and 
universal finite size amplitudes at criticality, {Phys.\ Rev.\ 
Lett.} {56} (1986) 742.

\bibitem{Carlipc} S.\ Carlip, What we don't know about BTZ black 
hole entropy, {Class.\ Quant.\  Grav.} {15} (1998)
3609, arXiv:hep-th/9806026, section 2.

\bibitem{Bala} V.\ Balasubramanian, J.\ de Boer, and D.\ Minic,
Mass, entropy and holography in asymptotically de Sitter spaces,
{Phys.\ Rev.} {D65} (2002) 123508, arXiv:hep-th/0110108.

\bibitem{Maloney} R.\ Bousso , A.\ Maloney, and A.\ Strominger,
Conformal vacua and entropy in de Sitter space, {Phys.\ Rev.} 
{D65} (2002) 104039, arXiv:hep-th/0112218.

\bibitem{Umetsu} H.\ Umetsu and N.\ Yokoi, Comments on 
quantum aspects of three-dimensional de Sitter gravity,
{Nucl.\ Phys.} {B650} (2003) 420, arXiv:hep-th/0208171.

\bibitem{Park} M.-I.\ Park, Statistical entropy of three-dimensional 
Kerr-de Sitter space, {Phys.\ Lett.} {B440} (1998) 275,
arXiv:hep-th/9806119.

\bibitem{BBO} M.\ Ba{\~n}ados, T.\ Brotz, and M.~E.\ Ortiz,
Quantum three-dimensional de Sitter space, {Phys.\ Rev.} 
{D59} (1999) 046002, arXiv:hep-th/9807216.

\bibitem{Kelly} W.~R.\ Kelly and D.\ Marolf, Phase spaces for 
asymptotically de Sitter cosmologies, {Class.\ Quant.\ Grav.} 
{29} (2012) 205013, arXiv:1202.5347. 

\bibitem{Parkx} M.-I.\ Park, BTZ black hole with gravitational 
Chern-Simons: thermodynamics and statistical entropy,
{Phys.\ Rev.} {D77} (2008) 026011, arXiv:hep-th/0608165.

\bibitem{Kaul} R.~K.\ Kaul and P.\ Majumdar, Schwarzschild 
horizon dynamics and SU(2) Chern-Simons theory, {Phys.\ Rev.} 
{D83} (2011) 024038, arXiv:1004.5487.

\bibitem{Deser} S.\ Deser and R.\ Jackiw, Three-dimensional 
cosmological gravity: dynamics of constant curvature,
{Annals Phys.} {153} (1984) 405.

\bibitem{sl} A.~F.\ Beardon, The geometry of discrete groups,
in \emph{Discrete groups and automorphic functions}, edited
by W.~J.\ Harvey (Academic Press, London, 1977).

\bibitem{Roken} C.\ R{\"o}ken, SL(2, C) and SU(2) connection 
variable formulations of Kerr isolated horizon geometries for 
loop quantum gravity, arXiv:1303.2548. 

\bibitem{GhoshPerez}  A.\ Ghosh and A.\ Perez, Black hole entropy and
isolated horizons thermodynamics, {Phys.\ Rev.\ Lett.} {27}
(2011) 241301, arXiv:1107.1320.

\bibitem{CEntropy} S.\ Carlip, Effective conformal descriptions of 
black hole entropy, {Entropy} 13 (2011) 1355, arXiv:1107.2678.

\bibitem{Pranzettib}  D.\ Pranzetti, personal communication, 2014.

\bibitem{Krasnov} K.\ Krasnov, 3D gravity, point particles and 
Liouville theory, {Class.\ Quant.\ Grav.} {18} (2001) 1291,
arXiv:hep-th/0008253.

\bibitem{Furstenberg} H. Furstenberg, Noncommuting random 
products, {Trans.\ Amer.\ Math.\ Soc.} {108} (1963) 377.

\bibitem{Quint} J.-F.\ Quint, personal communication, 2014.

\bibitem{Jacobson} T.\  Jacobson and R.\ Parentani, Horizon entropy,
   {Found.\ Phys.} {33} (2003) 323, arXiv:gr-qc/0302099.

\bibitem{Husain} V.\ Husain, Apparent horizons, black hole entropy 
   and loop quantum gravity, {Phys.\ Rev.} {D59} (1999) 084019,
   gr-qc/9806115.

\bibitem{KrasnovBF} K.\ Krasnov, Gravity as BF theory plus potential,
   {Int.\ J.\ Mod.\ Phys.} {A24} (2009) 2776, arXiv:0907.4064.

\end{thebibliography}
\end{document}